\documentclass[12pt]{article}
\usepackage{epsfig}

\begin{document}

\title{Non polynomial conservation law densities  generated
by the symmetry operators in some hydrodynamical models} 
\author{Maxim V. Pavlov,  Ziemowit Popowicz \\
 University of Wroc\l aw,  Institute of Theoretical Physics\\
 pl.M.Borna 9  50-205 Wroc\l aw Poland\\
 e-mail: ziemek@ift.uni.wroc.pl }
\maketitle

\begin{abstract}
New extra series of conserved densities for the polytropic gas model and nonlinear elasticity equation 
are obtained without 
any references to the recursion operator or to the Lax operator formalism. 
Our method based on the utilization of the symmetry operators and allows us to obtain the 
densities of arbitrary homogeneity dimensions. 
The non polynomial densities  with logarithmic behavior are presented as an example.
The special attention is paid for the singular case 
$(\gamma=1)$  for which we found new non homogeneous solutions expressed in terms of the elementary functions.
\end{abstract}

\section*{Introduction}

The conservation laws are the most important object in the classical mechanics 
as well as in the field theory. There are many different methods of the 
constructions of these laws. The most popular, especially used in the 
soliton theory and in the hydrodynamics,  utilize  the so called recursion 
operator or Lax operator formalism \cite{solo}. On the other hand, it appeared that for  the  Nonlinear Schroedinger 
equation it is possible to find one series of conserved Hamiltonians using 
recursion operator only. However for the shallow water equation,
which is the dispersionless limit of the Nonlinear Schroedinger equation, there 
is additional series  of conserved densities which  is impossible to obtain by 
recursion operator (see for example \cite{Pavlov}). 

    In this paper we would like to show that it is possible to construct 
new extra series of conserved densities for the polytropic gas model and nonlinear elasticity equation
 \cite{Nutek1,Das} 
avoiding using recursion operator  or Lax formalism. 
More precisely we generate many  nonequivalent Hamiltonians of the given dimensions, 
using symmetry operator. 
Our Hamiltonians are  non polynomial expressions  which contains  logarithmic functions.
Independently we consider the singular case, for the polytropic gas system for which $(\gamma =1)$ . 
For this system we constructed new series 
of  non homogeneous solutions expressed in terms of the elementary functions.

    The paper is organized as follows. In the first  section we describe the basic properties of the 
polytropic gas system which we use in the next sections. In the second section we describe our symmetry 
approach where we utilize the shift, scaling and  projective operators in order to generate the conserved 
densities. In the third section we present  explicitly new 
series of the conserved densities for different, physically interesting, models of the polytropic 
gas system with  $\gamma = 2,3,4,5,{5\over3},{7\over 5}, -1$. In the last section we adopt our
formalism to the degenerated case 
$(\gamma=1)$ which is obtained  from the contraction of the previous case. We show that in this case 
the conserved densities are  connected with the Bessel equation.

\section{The hydrodynamical systems.}
The theory of the hydrodynamical  type systems of the nonlinear equations \cite{Dubrov}
\begin{equation}
u_{t}^{i}=\sum_{i=1}^{N} \upsilon _{j}^{i}({\bf u})u_{x}^{j}
\quad i,j=1,2, . . . , N  \label{1}
\end{equation}
where $ {\bf u} = ( u_1,u_2, ... u_N) $ and $\upsilon_j^i$ are some functions ,  integrable by the generalized hodograph method 
\cite{Tsarev} is closely related
to the over-determined systems of first order partial differential linear equations.
The conservation laws are such that   
\begin{equation}
\frac{\partial h}{\partial t} = \frac{\partial g}{\partial x} \label{conserv}
\end{equation}
where $ h$ is  density and $g$ is flux. Then densities satisfy    
\begin{equation}
\partial _{k}(\partial _{i}h)=\Gamma _{ik}^{i}(\partial _{i}h)+\Gamma
_{ik}^{k}(\partial _{k}h). \nonumber\\
\quad i\neq k,  \label{2}
\end{equation}
where 
\begin{equation}
\Gamma_{ik}^{i}\equiv \frac{ \partial _{k}\mu ^{i}}{\mu ^{k}-\mu^{i}} \quad  i\neq k \quad 
\partial _{k}\equiv \partial /\partial r^{k} 
\end{equation} 
and $r^{k}({\bf u})$ are the so called Riemann invariants
in which the hydrodynamic type system (\ref{1}) is rewritten in the diagonal form 
\begin{equation}
r_{t}^{i}=\mu ^{i}(\mathbf{r})r_{x}^{i},  \label{3}
\end{equation}
and no summation on the repeated indices.
Thus, (\ref{2}) is a linear systems of {\it first} order partial differential equations 
with variable coefficients. The general solution of such system is  
determined up to  $N$ arbitrary functions of a single variable. 

There are many particular cases of the system (\ref{2})  for which 
a general solution is expressed in explicit and in  compact form
(\cite{Pavlov2}). 
If we cannot to find a general solution of such system, then alternative way to solve the  
Cauchy or Goursat problems is to create the  infinite number of particular 
solutions \cite{Tsarev1}. 

It appeared that the conserved densities can be used in the 
construction of particular solutions. Indeed. Let us consider  the polytropic gas 
\begin{equation}
\eta _{t}=\partial _{x}(u\eta ), \quad  u_{t}=\partial _{x}\Big [\frac{u^{2}
}{2}+\frac{\eta ^{\gamma -1}}{\gamma -1}\Big ],
\end{equation}
and nonlinear elasticity equations
\begin{equation}
\eta _{y}=u_{x} , \quad  u_{y}=\partial _{x}\Big [\frac{\eta ^{\gamma -2}}{
\gamma -2}\Big ],
\end{equation}
which are the commuting flows to each other.

The solution of the first system obtained by hodograph method is 
\begin{equation}
t = h -\eta\frac{\partial h}{\partial \eta},  \quad x = -\frac{\partial h}{\partial \eta} -
{u \over \eta} \frac{\partial h}{\partial u} \label{4}
\end{equation}
while for the second is 
\begin{equation}
x = \frac{\partial h}{\partial \eta} , \quad   
y = \frac{\partial h}{\partial u} \label{5}
\end{equation}
where $h$ is some solution of the Tricomi like  equation 
\begin{equation}
h_{uu}=\eta ^{3-\gamma }h_{\eta \eta }.  \label{6}
\end{equation}
The previous equation is nothing but the equation on the conservation law densities for
both systems.  Formulas (\ref{4}) and (\ref{5}) 
realize the general or particular solution for both systems if $h$ is the general or particular 
solution of the equation (\ref{6}) respectively. 

In \cite{Nutek1} and \cite{Das}
two infinite serieses of {\it quasi linear} conservation laws were
constructed for the polytropic gas and for the nonlinear elasticity equations respectively.
{ \it Quasi linear} means, that these conservation law
densities are polynomials with respect to $u$, $\eta $ and $\eta ^{\gamma}$ where $\gamma$ is an 
arbitrary polytropic constant. 

However, as we see in the next section,  these densities does not exhaust  the set of all 
possible conserved densities.

\section{Symmetry operator approach for $\gamma \neq 1$.}

Let us first consider the more general form of the polytropic gas system 
\begin{eqnarray}
\eta _{t} &=&\partial _{x}(u\eta ), \qquad u_{t}=\partial _{x}[\frac{u^{2}}{2}+
\eta f^{\prime \prime }(\eta )-f^{\prime }(\eta )], \\
\eta _{y} &=&u_{x}, \qquad \qquad u_{y}=\partial _{x}f^{\prime \prime}(\eta )
\end{eqnarray}
where $f$ is an arbitrary function. Our generalized system constitute  Hamiltonian equations with 
the following local structure
\begin{equation}
\eta _{t}=\partial _{x}\frac{\delta H}{\delta u}, \quad 
u_{t}=\partial _{x}\frac{\delta H}{\delta \eta }, \quad 
\eta_{y}=\partial _{x}\frac{\delta \tilde{H}}{\delta u}, \quad
u_{y}=\partial _{x}\frac{\delta \tilde{H}}{\delta \eta },
\end{equation}
where 
\begin{eqnarray}
H &=& \int [\frac{1}{2}u^{2}\eta +\eta f^{\prime }(\eta )-2f(\eta )]dx, \ \nonumber\\ 
\tilde{H} &=& \int [\frac{u^{2}}{2}+f^{\prime}(\eta )]dx.
\end{eqnarray}

In order to obtain the conserved densities 
we try to eliminate the differentials $dq$ and $dp$ from the corresponding conservation laws
\begin{equation}
\partial_{t}h=\partial _{x}p,\qquad  \partial _{y}h=\partial _{x}q
\end{equation}
By direct calculation we have 
\begin{equation}
dq=h_{\eta }du+f^{\prime \prime \prime }(\eta )h_{u}d\eta, \qquad 
dp=[uh_{u}+\eta h_{\eta }]du+[\eta f^{\prime \prime \prime }(\eta
)h_{u}+uh_{\eta }]d\eta 
\end{equation}
The compatibility conditions ( $ (p_u)_{\eta} =(p_{\eta})_u , \quad (q_u)_{\eta}=(q_{\eta})_u$ ) 
lead to the Tricomi - like equation
\begin{equation}
h_{\eta \eta }=f^{\prime \prime \prime }(\eta )h_{uu},
\end{equation}
Since, this equation is compatible with the shift symmetry operator
(see \cite{Pavlov3})
\begin{equation}
\delta =\partial /\partial u,
\end{equation}
one can search solutions in the form
\begin{equation}
h_{u}=\lambda h,
\end{equation}
where $\lambda $ is an arbitrary parameter.
This is the eigenvalue problem for the shift symmetry operator.
Then Tricomi-like equation is transformed
to the  linear ordinary differential equation
\begin{equation}
h_{\eta \eta }=\lambda ^{2}f^{\prime \prime \prime }(\eta )h.
\end{equation}
However this equation cannot be solved explicitly  for an arbitrary
$f$.

We propose, from that reason, quite different  approach. 
Let us observe that the shift symmetry operator transforms one solution of the 
Tricomi - like  equation onto another one
\begin{equation}
\partial _{u}h_{n+1}=h_{n}.
\end{equation}
It means that all conservation law densities $h_{k}$ and corresponding fluxes $p_{k}, q_{k}$
can be written down  in the quadratures recursively
\begin{eqnarray}
dh_{k+1} &=&h_{k}du+q_{k}d\eta,  \quad dq_{k+1}=q_{k}du+f^{\prime
\prime \prime }(\eta )h_{k}d\eta , \\
dp_{k+1} &=&[uh_{k}+\eta q_{k}]du+[\eta f^{\prime \prime \prime }(\eta
)h_{k}+uq_{k}]d\eta .
\end{eqnarray}
Interestingly all known examples of the conserved densities could be obtained in this way.

However we demonstrate different possibilities which give us new serieses of the conserved densities. 
Our main idea is based on the following observation. 

The Tricomi like equation (\ref{6}) is compatible with three  local
symmetry operators:
\vspace{0.3cm}

1. Shift operator
\begin{equation}
\delta =\frac{\partial }{\partial u},
\end{equation}

2. Scaling operator
\begin{equation}
R=u\frac{\partial }{\partial u}+\frac{2}{\gamma -1}\eta \frac{\partial }
{\partial \eta }, \label{scala}
\end{equation}

3. Projective operator
\begin{equation}
S=[\frac{\gamma -1}{4}u^{2}+(\gamma -1)^{-1}\eta ^{\gamma -1}]\frac{\partial }{
\partial u}+u\eta \frac{\partial }{\partial \eta }+\frac{\gamma -3}{4}u. \label{algebra}
\end{equation}
These operators, with the identity operator, constitute  closed Lie Algebra, with 
the following commutation relation 
\begin{equation}
\big [ \delta ,S \big ] = {2\over \gamma-1 } R + {\gamma-3\over 4}, \qquad 
\big [ \delta ,R \big ] = \delta,  \qquad
\big [ R ,S  \big ] = S \label{com}
\end{equation} 
They act on {\it homogeneous} conservation law densities as follows
\begin{equation}
\delta h_{k+1}=h_{k}\quad Rh_{k}=c_{k}h_{k}\quad 
Sh_{k}=h_{k+1},\label{7}
\end{equation}
where $c_{k}$ are degree of homogeneity. Thus, by combination of these
symmetry operators one can describe all quasi linear conservation law
densities.

Interestingly if we rewrite the Tricomi -like equation using the Riemann invariants 
we obtain the famous Euler - Darboux - Poisson equation
\begin{equation}
\frac{\partial h}{\partial r_{1}\partial r_{2}}=\frac{\varepsilon }{
r_{1}-r_{2}}\Big [\frac{\partial h}{\partial r_{1}}-\frac{\partial h}{\partial
r_{2}}\Big ], \label{tricom1}
\end{equation}
where
\begin{equation}
\varepsilon =\frac{3-\gamma}{2(1-\gamma)}, \quad r_{1}=u+{2\over(\gamma-1)}\eta^{\frac{\gamma-1}{2}},
\quad r_{2}=u-{2\over(\gamma-1)}\eta^{\frac{\gamma-1}{2}}.
\end{equation}
Now the symmetry operators are   
\begin{eqnarray}
\delta  &=&\frac{\partial }{\partial r^{1}}+\frac{\partial }{\partial
r^{2}} ,  \quad 
R=r^{1}\frac{\partial }{\partial r^{1}}+r^{2}
\frac{\partial }{\partial r^{2}},  \label{sim} \\
S &=&(r^{1})^{2}\frac{\partial }{\partial
r^{1}}+(r^{2})^{2}\frac{\partial }{\partial r^{2}}+\varepsilon \lbrack r^{1}+r^{2}].
\end{eqnarray}
The "zero--solutions" of the shift operator $\delta $
\begin{equation}
\delta h_{0}=0
\end{equation}
can be found immediately
\begin{equation}
h_{0}^{(1)}=1,  \quad h_{0}^{(2)}=\eta,
\end{equation}
while for the projective operator one can obtain
\begin{equation}
h_{1}^{(1)}=u, \quad h_{1}^{(2)}=u\eta ,
\end{equation}
etc. 

The "zero--solutions" of projective operator are easy to obtain also from Riemann
invariant ($r^{1}$, $r^{2}$) form
\begin{equation}
h=(r^{1}r^{2})^{-\varepsilon }.
\end{equation}
If we shift the Riemann invariants, in this formula, on  arbitrary parameter $\lambda$ we obtain 
the generating function on the  conservation law densities. When $\lambda \rightarrow \infty$ 
we can obtain known densities.

In order to obtain new densities we demonstrate quite different possibilities .
Let us consider the following recursive chain 
\begin{equation}
Rh_{k}^{(0)}=c_{k}h_{k}^{(0)}, \quad 
Rh_{k}^{(1)}=c_{k}[h_{k}^{(1)}+h_{k}^{(0)}], \quad 
Rh_{k}^{(2)}=c_{k}[h_{k}^{(2)}+h_{k}^{(1)}] \label{8}
\end{equation}
etc..., where $h_{k}^{(0)}$ are the quasi linear  conserved densities.  

Now $h_{k}^{(i)}, i=1,2,...$ are new conserved densities which  satisfy the 
Tricomi-like equation (\ref{6}).
Notice that the second formula in   (\ref{7}) is not in  contradiction  to  previous chain because 
formula (\ref{7}) is valid for homogeneous  conservation laws only.  

Now let us consider the shallow water equation, e.g. the polytropic gas system 
with $\gamma = 2$. Let us choose the first homogeneous solutions of (\ref{7}) 
\begin{equation}
h^{(0)}_1=u  , \qquad h^{(0)}_2=\eta, \qquad h^{(0)}_3 = u\eta \qquad h^{(0)}_4=u^2\eta + \eta^2 
\end{equation}
etc. 
As the result we obtained that 
\begin{eqnarray}
h^{(1)}_1 &=& 2 {\sqrt {u^2 - 4\eta}}\Big ( \ln  \Big ({\sqrt { u^2 - 4\eta }}+u\Big )  - \ln 2 \Big ) +
 \Big (u - {\sqrt { u^2 - 4\eta }}  \Big ) \ln \eta \ \nonumber\\ 
h^{(1)}_2 &=&  \frac{u^2}{2} +\eta \ln \eta  \ \nonumber\\
h^{(1)}_3 &=& {1\over 4} \Big ( u^3 + 6u\eta \ln \eta \Big ) \ \nonumber\\
h^{(1)}_4 &=& {1\over 6} \Big ( u^4+12u^2\eta \ln \eta + 12\eta^2 \ln \eta -18\eta^2 \Big )
\end{eqnarray}
are  conserved density also. In the next let us apply this procedure to the $h^{(1)}_2$ and $h^{(1)}_3$ 
\begin{eqnarray}
h^{(2)}_2 &=& {1\over 18}\Big ( -36\eta \ln ^2({ \sqrt{z}}+u) + 18 \ln ({\sqrt{z}}+u)
\big ( 2\eta u \ln (4\eta) -2\eta + u{\sqrt{z}} \big ) \ \nonumber \\
&& +9u{\sqrt{z}}(1-\ln (4\eta )) -36\eta \ln \eta\ln 2 +\big (9u^2+18\eta \big ) \ln \eta -18u^2+32\eta 
\Big ), \ \nonumber\\  
h^{(2)}_3 &=&  {1\over 12}\Big ( -36\eta u \ln ^2({ \sqrt{z}}+u) + \ln ({\sqrt{z}}+u)
\big( 36 \eta u \ln (4\eta)-6\eta u + \ \nonumber\\ 
&& 6{\sqrt{z}}(u^2+8\eta) \big ) -36\eta u \ln(\eta) \ln 2 
 +3\ln \eta \big (u^3 +6\eta u - {\sqrt{z}}(u^2+8\eta) \big ) +
\ \nonumber\\
&& 3{\sqrt{z}} \big(1 -2\ln 2 \big)  \big (u^2+8\eta \big ) -16\eta u-8u^3 \Big ) ,
\end{eqnarray}
where $z=u^2-4\eta$. 

We obtained rather complicated formulas which contain the logarithm 
in the second power. Now we can continue this procedure obtaining complicated formulas which contain 
the logarithm in higher powers. On the other side we can combine our method with this which generate 
quasi linear densities obtaining following diagram
\newpage

\begin{figure}[tbh]                           
\begin{center}                              
{\includegraphics[width=10cm]{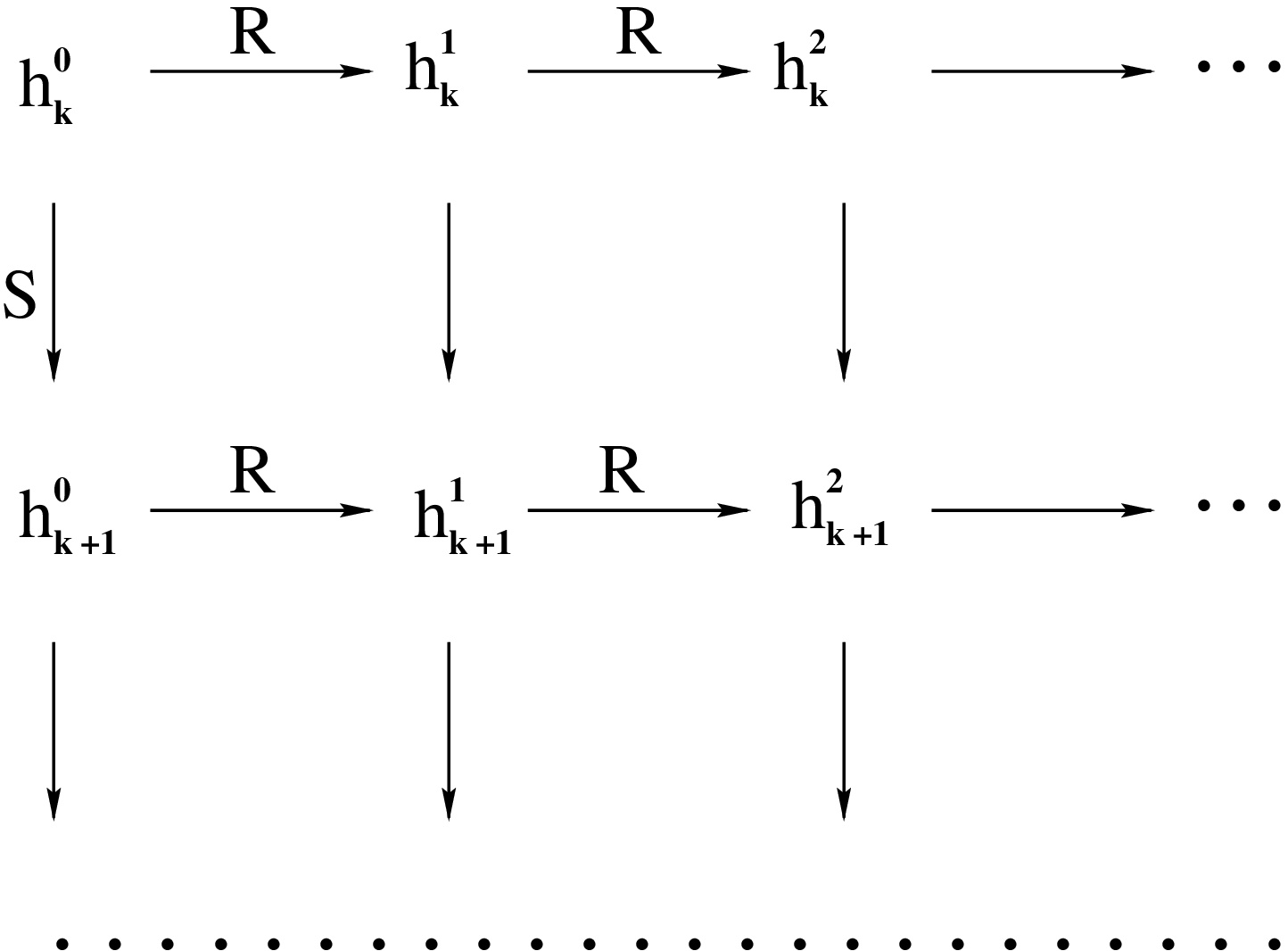}} 
\end{center}                                
\end{figure}

This diagram obviously is closed. In order to see it let us start consideration from some $h^{(n)}_k$. 
Applying at first projective and at next step scaling transformations we obtain 
\begin{equation}
R  h^{(n+1)}_{k+1} =c_{k+1}\Big ( h^{(n+1)}_{k+1} + Sh^{(n)}_k \Big )
\end{equation}  
Finally applying these operators in the reverse order we obtain 
\begin{equation}
S R  h^{(n+1)}_{k} =c_{k}\Big (Sh^{(n+1)}_{k} + Sh^{(n)}_k\Big )
\end{equation}  
Using the commutation relation (\ref{com}) we see that these formulas coincide because $c_{k+1} = c_k +1$.

\section{Non polynomial densities for an arbitrary $\gamma \neq 1$ in polytropic gas system}

For an arbitrary $\gamma$ our generating equations become
\begin{equation}
R h^{(n)}_k = c_{k} \Big ( h^{(n)}_{k} + h^{(n-1)}_{k} \Big ). \label{general}
\end{equation}
Now we have to solve the Tricomi like equation  on $h^{(n)}_k$ which is however rather hard to do. 
We restrict, from that reason, to two cases where  $ h^{(0)}_1 = u $ and $h^{(0)}_2=\eta$.

For the first case we have $c_1=1$. By direct verification one can check that 
\begin{equation}
h^{(1)}_1=uh(\tau) + u \ln u
\end{equation}
where $\tau={u^2\over \eta^{\gamma-1}}$ satisfy (\ref{general}). The Tricomi like equation reduces 
in this case to 
\begin{equation}
\tau^2 \Big ( 4-(\gamma-1)^2 \Big )\frac{\partial^2 h}{\partial^2 \tau} + \Big (6-\gamma (\gamma -1)\tau
\Big ) \frac{\partial h}{\partial \tau} +1 = 0.
\end{equation}

For the second case we have $c_2={2\over \gamma-1}$. Similarly to the first case,  by direct 
verification, one can check that 
\begin{eqnarray}
h^{(1)}_2 &=& \eta h(\tau) + \eta \ln \eta  \ \nonumber\\
\end{eqnarray}
satisfy (\ref{general}).
Substituting  $h^{(1)}_2$ to the Tricomi like equation (\ref{6}) we obtain the following equation on 
the function $h(\tau)$
\begin{equation}
 \tau \Big ( 4 -\tau (\gamma -1)\Big ) \frac{\partial^2 h}{\partial^2 \tau} -
 \Big (2 - (\gamma-1)(\gamma-2)\tau \Big ) \frac{\partial h}{\partial \tau} + 1 =0
\end{equation}
In both cases it is possible to obtain closed formulas for particular values of $\gamma$ parameter.
However we can simplify our consideration using different coordinates. For example choosing 
the  coordinates  $r$ and $p$ as \cite{Nutek1}
\begin{equation}
\eta=rp, \qquad u={1\over \gamma-1}\Big ( r^{\gamma-1} +p^{\gamma-1} \Big ),
\end{equation} 
we can rewrite the scaling symmetry operator and Tricomi like equation as
\begin{eqnarray}
R &=& {1\over  \gamma-1} \Big (r\frac{\partial }{\partial r} +p\frac{\partial }{\partial p} \Big ) \ \nonumber \\
 p^{3-\gamma}\frac{\partial^2 h}{\partial^2 p} &=& r^{3-\gamma}\frac{\partial^2 h}{\partial^2 r} 
\end{eqnarray}
Taking into an account  $h^{(0)}_1=r$ we obtained  the following solutions on the $h^{(1)}_1$ 
densities
\begin{eqnarray}
\gamma=2,  \qquad h^{(1)}_1 &=& - (r-p)\ln(r-p) +p\ln(p) \ \nonumber\\
\gamma=3,  \qquad h^{(1)}_1 &=&  (r+p)\ln(r+p) - (r-p)\ln(r -p) \ \nonumber\\
\gamma=4,  \qquad h^{(1)}_1 &=& -(r-p)\ln(r-p) +(r+2p)\ln(p^2+pr+r^2) \ \nonumber\\ 
&& \qquad  + {\sqrt{ 3 }} p \arctan ( \frac{p+2r}{{ \sqrt{ 3 }}p}) \ \nonumber\\
\gamma=5,  \qquad h^{(1)}_1 &=& (p+r)\ln (p+r) + (r-p)\ln(r-p) + r\ln(p^2+r^2) \ \nonumber\\
&& \qquad  + 2p \arctan(\frac{r}{p}) 
\end{eqnarray}
Finally let us present the solution for $\gamma ={5\over 3},{7\over 5}$  and $\gamma=-1$ which are 
interesting from the physical point because these describe dynamics of one-atomic  gas and 
two-atomic gas  \cite{atom} and two dimensional nonlinear Born-Infeld electrodynamics \cite{borninf}
respectively.
\begin{eqnarray}
\gamma={5\over 3},&&  \qquad h^{(1)}_1 = \frac{3}{4} p{~}  {\rm arctanh}\tau^{\frac{1}{3}} + \frac{r}{2}
\ln \big( 3\tau^{\frac{2}{3}} - 3\tau^{\frac{4}{3}} + \tau^2 -1 \big ) 
\ \nonumber\\
&& \qquad \qquad \qquad - r \ln \tau -3r\tau^{-\frac{2}{3}} \ \nonumber\\
\gamma={7\over 5}, && \qquad h^{(1)}_1 =  \frac{5}{4} p {~}  {\rm arctanh}\tau^{\frac{1}{5}} + \frac{r}{2}
\ln \big(-10 \tau^{\frac{4}{5}} - 5\tau^{\frac{8}{5}} + 
5\tau^{\frac{2}{5}}+\tau^{\frac{6}{5}}+\tau^2 -1 \big ) \ \nonumber\\
&& \qquad \qquad \qquad  - r \ln \tau -\frac{5}{3} r\tau^{-\frac{2}{5}} - 5 r\tau^{-\frac{4}{5}} + r\ln r \ \nonumber\\
\gamma=-1, && \qquad  h^{(1)}_1 = {1\over 2} \Big ( r \ln \big( {p^2r^2\over r^2-p^2} \big )  + p\ln 
\big ( {r-p\over r+p} \big )\Big )
\end{eqnarray}

\section{ Symmetry operator approach for $\gamma = 1$ and corresponding  non polynomial densities.}

For this case the polytropic gas  and the nonlinear elasticity systems  are 
\begin{equation}
\eta _{t}=\partial _{x}(u\eta ), \quad  u_{t}=\partial _{x}\Big [\frac{u^{2}
}{2}+\ln \eta \Big ].
\end{equation}
\begin{equation}
\eta _{y}=u_{x} , \quad  u_{y}=\partial _{x} \big (-\frac{1}{\eta } \big ),
\end{equation}
respectively.
For this case  the Euler - Darboux - Poisson equation (\ref{tricom1}) 
degenerates because $\varepsilon \rightarrow \infty$. Similarly the scaling operator $ R $ 
(see (\ref{scala})) also becomes degenerated.

In order to solve this problem we notice that the analogue of the symmetry algebra (\ref{algebra}) 
can be obtained considering the contraction  
with respect to  $\gamma = 1$.  If we rescale $R \rightarrow \frac{\gamma-1}{2} R$ 
and compute the limit when $\gamma \rightarrow 1$ we obtain
\begin{equation}
\delta =\partial _{u}, \quad R=\eta \partial _{\eta }, \quad 
S=\ln \eta \partial _{u}+u\eta \partial _{\eta }-u/2, \label{symetr}
\end{equation}

Interestingly now, in the contraction limit,  the homogeneity properties 
(\ref{7}) does not hold. It means that the Tricomi-like equation
\begin{equation}
h_{uu}=\eta ^{2}h_{\eta \eta }
\end{equation}%
has no  any homogeneous solutions.  

However the quasi-homogeneous solutions could be obtained (\cite{Nutek2}).
By quasi-homogeneous  we understand the homogeneous with respect to the one variable 
$ u $ or $\ln \eta$.

For our further applications 
we describe  another possibility.  We constructed the generating function for the conserved densities. 
To end this we consider the eigenvalue problem for each symmetry operator.

1. For the scaling operator $ R $ (see (\ref{symetr})) the eigenvalue problem
\begin{equation}
h_{\xi }=\lambda h \label{sim2},
\end{equation}
where $\partial _{\xi }\equiv \eta \partial _{\eta }$,
reduces the Tricomi-like equation to the linear ordinary differential
equation of the second order
\begin{equation}
h_{uu}=\lambda (\lambda -1)h. 
\end{equation}

These equations can be easily integrated and we obtained 
\begin{equation}
h=\exp [\sqrt{\lambda (\lambda -1)}u+\lambda \xi ]. \label{trykom}
\end{equation}

2. For the shift operator $\delta$ (see (\ref{symetr})) the eigenvalue problem 
\begin{equation}
h_{u}=\tilde{\lambda}h \label{sim1}
\end{equation}
reduces the Tricomi-like equation to the linear ordinary differential
equation of the second order
\begin{equation}
h_{\xi \xi }-h_{\xi }=\tilde{\lambda}^{2}h,
\end{equation}
These equation are  easily also to integrate and the solutions have the same form as (\ref{trykom})
in which we should replace $\lambda^2 \rightarrow  \tilde{\lambda} (\tilde{\lambda}-1)$.

If one expands function $h$ near $\lambda \rightarrow 0$ or $\lambda
\rightarrow 1$, then one can obtain two well-known infinite serieses of
conservation law densities (\cite{Nutek2}).

3. Projective operator $ S $ (see (\ref{symetr})). If one substitutes
\begin{equation}
h=q\exp [\xi /2], \quad u=2s\cosh \theta,  \quad \xi
=2s\sinh \theta ,
\end{equation}
where $s$ and $\theta$ are new functions, then the eigenvalue problem looks like  
\begin{equation}
q_{\theta }=\lambda q
\end{equation}
Now the generating function is 
\begin{equation}
q=\psi (s)\exp [\lambda \theta ],
\end{equation}
where the function  $\psi$ satisfies the Bessel equation
\begin{equation}
 \psi^{\prime \prime}  +{1\over s}\psi^{\prime}+ [1-{\lambda ^{2}\over s^2} ] \psi  =0 ,
\end{equation}

Let us present the most simplest example of the conserved densities obtained in this way, where 
it is possible to describe the Bessel function by the elementary functions for which 
$\lambda=\frac{1}{2}$
\begin{equation}
h={\sqrt {{\eta \over u-\ln \eta}}} \cos \Big ({{\sqrt{ u^2-\ln^2 \eta }}\over 2} \Big ).
\end{equation}
Using the recursion relations, known in the theory of Bessel functions, it is possible to 
generate infinite sets of conserved densities written in terms of elementary functions without any 
references to the recursion operator,  which appear in the Hamiltonian  approach to the polytropic 
gas systems. Finally we can use the shift or scaling symmetry operators and generate 
some conserved densities as we did in the previous sections.

\section{Conclusion}
In this paper we constructed 
new extra series of conserved densities for the polytropic gas model and nonlinear elasticity equation
avoiding using recursion operator  or Lax formalism. 
Our Hamiltonians  appeared as the  non polynomial  expressions  which contain  logarithmic functions.
If we continue our procedure to these logarithmic densities in the next step we obtain expressions with 
the logarithmic function in an arbitrary power.
We  considered the singular case of the polytropic gas system also, for which we found 
the  non homogeneous solutions expressed in terms of Bessel functions.
If we apply the point transformation to the symmetry operators then they can change the role. It means that
the scaling  and shift operators are equivalent to each other, what it easy to see in the 
Riemann invariants (\ref{sim}) or (\ref{sim1},\ref{sim2}). 
We presented this method for two hydrodynamical 
systems only, but this method can be adopted to more complicated equations as well.

\section*{Acknowledgments} One of the author (M.P.) thanks the Institute of Theoretical Physics University of 
Wroc\l aw for their hospitality.


\begin{thebibliography}{9}

\bibitem{solo} M.B\l aszak ,,{\it Multi-Hamiltonian Theory of Dynamical Systems} " Springer -Verlag 1998,
P.Olver ,, {\it Applications of Lie Group to Differential Equations} " ,
Springer - Verlag 1989. 

\bibitem{Pavlov} M. V. Pavlov, S. P. Tsarev, Uspekhi Mat. Nauk, 46, No.
4, 169--170 (1991).

\bibitem{Nutek1} Y.Nutku, M.Pavlov J.Math.Phys {\bf 43} (2002) 1441.

\bibitem{Das} J.C.Brunelli, A.Das Phys.Lett {\bf B} 426 (1998) 57.

\bibitem{Dubrov} B. A. Dubrovin, S. P. Novikov,  Dokl. Akad. Nauk
SSSR, 270, No. 4, (1993) 781.

\bibitem{Tsarev} S. P.Tsarev, 
Dokl. Akad. Nauk SSSR, 282, No. 3, 534--537 (1985).


\bibitem{Pavlov2} M.Pavlov ,, {\it Integrable hydrodynamic chains }" nlin.SI/00301010.

\bibitem{Tsarev1} S.P.Tsarev  Izv. Akad. Nauk SSSR, ser. Matem., 54, No. 5, 1048--1068 (1990).

\bibitem{Nutek2} Y.Nutku J.Math.Phys {\bf 28} (1987) 2579.


\bibitem{Pavlov3} M. V. Pavlov, S.P. Tsarev Funct. Anal.Appl {\bf 37} No.1 (2003) 38.

\bibitem{atom} B.L.Rozhdestvenski, N.N. Yanenko ,, {\it Systems of quasilinear equations and their 
applications to gas dynamics} " Mathematical Monographs, {\bf 55} AMS, Providence, RI (1983).

\bibitem{borninf} M.Arik, F.Neyzi, Y.Nutku, P.J.Olver, J.Verosky J.Math.Phys {\bf 30} (1988) 1338.
\end{thebibliography}
\end{document}